# Analysis and visualization of spatial transcriptomic data

Boxiang Liu[1][*][†], Yanjun Li[1][†], Liang Zhang[1]

[1]Baidu Research, Sunnyvale, CA, United States

[†]These authors have contributed equally to this work

[*]Correspondence: jollier.liu@gmail.com

## Abstract

Human and animal tissues consist of heterogeneous cell types that organize and interact in highly structured manners. Bulk and single-cell sequencing technologies remove cells from their original microenvironments, resulting in a loss of spatial information. Spatial transcriptomics is a recent technological innovation that measures transcriptomic information while preserving spatial information. Spatial transcriptomic data can be generated in several ways. RNA molecules are measured by *in situ* sequencing, *in situ* hybridization, or spatial barcoding to recover original spatial coordinates. The inclusion of spatial information expands the range of possibilities for analysis and visualization, and spurred the development of numerous novel methods. In this review, we summarize the core concepts of spatial genomics technology and provide a comprehensive review of current analysis and visualization methods for spatial transcriptomics.

## 1    Introduction

Quantification of gene expression has important applications across various aspects of biology. Understanding the spatial distribution of gene expression has helped to answer fundamental questions in developmental biology (1, 2), pathology (3, 4), cancer microenvironment (5-8), and neuroscience (9-11). Two widely used methods for gene expression quantification are fluorescent *in situ* hybridization (FISH) and next-generation sequencing. With FISH, fluorescently-labeled RNA sequences are used as probes to identify its naturally occurring complementary sequence in cells while preserving the spatial location of the target sequences (12). Traditionally, the number of target sequences simultaneously identified by *in situ* hybridization is restricted by the number of fluorescent channels, making this method suitable for targeted gene detection. On the other hand, next-generation sequencing methods use a shotgun approach to quantify RNA molecules across the entire transcriptome (13). To achieve transcriptome-wide quantification, RNA must be first isolated and purified, which removes RNA molecules from their native microenvironment. Even with single-cell sequencing, where the cellular origin of RNA molecules is preserved, spatial information of cells can only be inferred but not directly measured (14, 15).

Various approaches have been made to measure gene expression while preserving spatial information. Tomo-seq applied the principle of tomography to measure spatial transcriptomic information in 3D. In tomo-seq, tissue samples are sliced by cryosection and measured with RNA-seq. Each measurement corresponds to the average gene expression within a slice. Measurements are taken along multiple axes to reconstruct pixel-wise 3D gene expression information. (16) LCM-seq isolates single cells with laser capture microscopy (LCM) and measures captured cells with single-cell RNA-sequencing. LCM can capture cells of desired types and with specific spatial locations of the tissue specimen (17). While these methods retain the spatial location of RNA-seq measurements, they suffer from high labor costs and incomplete spatial coverage. In this review, we cover recent advances in spatial transcriptomic methods that attempt to address these challenges. In addition, we provide a comprehensive review of analysis and visualization techniques for spatial transcriptomic datasets.

The following sections are organized as follows (Figure 1). Section 2 discusses the latest developments in experimental spatial transcriptomic technologies. Section 3 discusses preprocessing of spatial transcriptomic data, an essential step prior to any analysis or visualization. Section 4 dissects methods whose inputs are gene expression without spatial coordinates. This includes dimensionality reduction, clustering, and cell-type identification. Section 5 describes methods whose inputs are gene expression combined with spatial coordinates. This includes identification of spatially coherent gene expression patterns and identification of spatial domains. Section 6 describes methods that analyze the interaction between cells or genes. All methods reviewed are listed in Table 1. This includes the identification of cell-to-cell communication and gene interaction. We note that other reviews on spatial transcriptomic technology(18-20) have been published during the peer review of this article.

**Table 1:** Current analysis and visualization tools for spatial transcriptomic datasets (accession date: 12/22/2021)

| Task | Tool | Inputs | Description | Language | Availability |
|---|---|---|---|---|---|
| **Preprocessing** | Space Ranger | Microscope images and FASTQ files | Space Ranger is an analysis pipeline for alignment, tissue and fiducial detection, barcode/UMI counting, and feature-spot matrix generation. | Bash and GUI | https://support.10xgenomics.com/spatial-gene-expression/software/pipelines/latest/what-is-space-ranger |
| | Scran (2016) (21) | Gene expression | Scran uses pool-based and deconvoluted cell-based size factors for single-cell gene expression normalization. | R | http://bioconductor.org/packages/release/bioc/html/scran.html |
| | SCNorm (2017) (22) | Gene expression | SCNorm uses double quantile regression-based model for gene-group normalization. | R | https://www.bioconductor.org/packages/release/bioc/html/SCnorm.html |
| **Clustering** | K-means | Gene expression | K-means iteratively assigns observations to the cluster with the nearest center. | R and Python | R: https://www.rdocumentation.org/packages/stats/versions/3.6.2/topics/kmeans; Python: https://scikit-learn.org/stable/modules/generated/sklearn.cluster.KMeans.html |
| | Gaussian mixture model | Gene expression | GMM is similar to K-means but softly assigns observations to clusters based on the Gaussian distribution. | R and Python | R: https://cran.r-project.org/web/packages/ClusterR/vignettes/the_clusterR_package.html; Python: https://scikit-learn.org/stable/modules/mixture.html |
| | hierarchical clustering | Gene expression | Hierarchical clustering iteratively merges closest observations. | R and Python | R: https://www.rdocumentation.org/packages/stats/versions/3.6.2/topics/hclust; Python: https://scikit-learn.org/stable/modules/clustering.html#hierarchical-clustering |
| | Louvain (2008) (23) | Gene expression | Louvain performs community detection within networks by iterative optimization of modularity. | R and Python | R: https://igraph.org/r/doc/cluster_louvain.html; Python: https://github.com/vtraag/louvain-igraph |
| | Leiden (2019) (24) | Gene expression | Leiden is a variant of the Louvain algorithm that guarantees well-connected communities. | R and Python | R: https://cran.r-project.org/web/packages/leiden/; Python: https://github.com/vtraag/leidenalg |
| | SC3 (2017) (25) | Gene expression | SC3 performs consensus clustering of single-cell RNA-seq data. | R | http://bioconductor.org/packages/release/bioc/html/SC3.html |
| | SIMLR (2017) (26) | Gene expression | SIMLR is a multi-kernel learning approach for single-cell RNA-seq clustering. | R and MATLAB | MATLAB: https://github.com/BatzoglouLabSU/SIMLR; R: https://www.bioconductor.org/packages/release/bioc/html/SIMLR.html |
| **Cell-specific marker genes** | scran (2016) (21) | Gene expression | Scran identifies consistently up-regulated genes through pairwise comparisons between clusters. | R | https://bioconductor.org/packages/devel/bioc/vignettes/scran/inst/doc/scran.html#6_Identifying_marker_genes |
| | scGeneFit (2021) (27) | Gene expression | ScGeneFit is a label-aware compressive classification method to select informative marker genes. | Python | https://github.com/solevillar/scGeneFit-python |
| **Cell-type identification** | scmap (2018) (28) | Gene expression | Scmap projects single-cell to reference data sets with an approximate k-nearest-neighbor search. | R | http://bioconductor.org/packages/release/bioc/html/scmap.html; Web version: https://www.sanger.ac.uk/tool/scmap/ |
| | SingleR (2019) (29) | Gene expression | SingleR iteratively calculates pairwise correlation across single cells and remove lowly correlated cell type for noise control. | R | https://github.com/dviraran/SingleR |
| | Cell-ID (2021) (30) | Gene expression of reference and target single-cell datasets | Cell-ID performs multiple correspondence analysis (MCA) based gene signature extraction and cell identification | R | https://bioconductor.org/packages/devel/bioc/html/CelliD.html |
| | JSTA (2021) (31) | *in situ* hybridization dataset | JSTA is a deep-learning-based cell segmentation and type annotation method by iteratively adjusting the assignment of boundary pixels based on the cell type probabilities for each pixel. | Python | https://github.com/wollmanlab/JSTA; https://github.com/wollmanlab/PySpots |
| **Dimensionality reduction** | Principal component analysis | Gene expression | PCA identifies orthogonal vectors that maximize the variance of projections from data points. | R and Python | R: https://www.rdocumentation.org/packages/stats/versions/3.6.2/topics/prcomp; Python: https://scikit-learn.org/stable/modules/generated/sklearn.decomposition.PCA.html |
| | t-SNE (2008) (32) | Gene expression | T-SNE iteratively refines projections in the low dimensional space to match pairwise distances in the high dimension space. | R and Python | R: https://cran.r-project.org/web/packages/Rtsne/; Python: https://scikit-learn.org/stable/modules/generated/sklearn.manifold.TSNE.html |
| | UMAP (2018) (33) | Gene expression | UMAP is similar to t-SNE but faster and better preserves high dimensional structure. | R and Python | R: https://cran.r-project.org/web/packages/umap/index.html; Python: https://umap-learn.readthedocs.io/en/latest/ |
| **Spatially coherent genes** | SpatialDE (2018) (34) | Gene expression + spatial coordinates | SpatialDE uses gaussian process regression to decompose variability into spatial and non-spatial components. | Python | https://github.com/Teichlab/SpatialDE |
| | Trendsceek (2018) (35) | Gene expression + spatial coordinates | Trendsceek uses marked point processes to identify spatial expression patterns. | R | https://github.com/edsgard/trendsceek |
| | Spark (2018) (36) | Gene expression + spatial coordinates | Spark is a generalized linear spatial model to identify spatial expression patterns. | R | https://xzhoulab.github.io/SPARK/ |
| **Spatial domains** | Zhu *et al.* (2018) (37) | Gene expression + spatial coordinates | Zhu *et al.* uses a hidden Markov random field to compare gene expression of neighboring cells to identify coherent expression patterns. | R and Python | R: https://bitbucket.org/qzhudfci/smfishhmrf-r/src/master/; Python: https://bitbucket.org/qzhudfci/smfishhmrf-py/src/master/ |
| | SpaGCN (2021) (38) | Gene expression + spatial coordinates + histology image | SpaGCN is a graph-convolutional-network-based method to jointly identify spatial domains and spatially variable genes. | Python | https://github.com/jianhuupenn/SpaGCN |
| **Spot deconvolution** | DSTG (2021) (39) | Gene expression + spatial coordinates | DSTG builds a graph consisting of real and pseudo spatial transcriptomic data and apply graph convolutional network to predict real data's cell type composition with help from pseudo data's label. | Python | https://github.com/Su-informatics-lab/DSTG |
| **Super-resolution** | BayesSpace (2021) (40) | Gene expression + spatial coordinates | BayesSpace is a Bayesian model to leverage neighborhood information to enhance resolution. | R | http://www.bioconductor.org/packages/release/bioc/html/BayesSpace.html |
| **Cell-cell interaction** | SpaOTsc (2020) (41) | Gene expression + spatial coordinates | SpaOTsc uses structured optimal transport between distribution of sender and receiver cells to identify cell-cell communication. | Python | https://github.com/zcang/SpaOTsc |
| **Receptor-ligand interaction** | GCNG (2020) (42) | Gene expression + spatial coordinates | GCNG is a graph convolutional neural network to encode the spatial information as a graph and to predict whether a gene pair will interact. | Python | https://github.com/xiaoyeye/GCNG |
| **Integrative** | Seurat (2018) (43) | Gene expression + spatial coordinates | Seurat is an R package for integrative single-cell transcriptomic analysis. | R | https://cran.r-project.org/web/packages/Seurat/index.html |
| | Giotto (2021) (44) | Gene expression + spatial coordinates | Giotto is an R package for integrative spatial transcriptomic analysis. | R | https://rubd.github.io/Giotto_site/ |
| | Scanpy (2018) (45) | Gene expression + | Scanpy is a Python package for integrative single-cell transcriptomic analysis. | Python | https://scanpy.readthedocs.io/en/latest/ |

| | | | | |
|---|---|---|---|---|
| | | spatial coordinates | | |
| | Squidpy (2021) (46) | Gene expression + spatial coordinates | Squidpy is a Python package for integrative spatial transcriptomic analysis. | Python | https://squidpy.readthedocs.io/en/stable/ |

## 2    Spatial transcriptomic technologies

Integration of spatial information with transcriptome-wide quantification has given rise to the emerging field of spatial transcriptomics. Currently, spatial transcriptome quantification falls into three broad categories (Table 2). First, spatial barcoding methods ligate oligonucleotide barcodes with known spatial locations to RNA molecules prior to sequencing (47-53). Both barcodes and RNA molecules are jointly sequenced, and spatial information of sequenced RNA molecules can be recovered from associated barcodes. Second, *in situ* hybridization methods coupled with combinatorial indexing can vastly increase the number of RNA species identified (54-57). The latest *in situ* hybridization methods can detect around 10,000 RNA species from a given sample (54). Third, *in situ* sequencing method uses fluorescent-based direct sequencing to read out base pair information from RNA molecules in their original spatial location (58, 59).

Several metrics need to be considered when selecting a method for a specific application (Table 2). Methods employing *in situ* hybridization provide subcellular resolution. Leveraging super-resolution microscopy, *in situ* hybridization methods can achieve a resolution of ~10nm, sufficient to distinguish single RNA molecules (60). In addition, *in situ* methods require no PCR amplification of cDNA, thus avoiding amplification bias. However, the number of RNA species detected by *in situ* methods is limited by the indexing scheme. The current detection limit is ~10,000 genes but will likely improve in the future. Furthermore, the area examined by *in situ* methods is limited by the field-of-view of the microscope objective lens. In contrast, spatial barcoding followed by shotgun sequencing can in principle sample the whole transcriptome. This is ideal if the target molecules are unknown *a priori*. Spatial barcoding can also examine larger tissue areas, making it ideal for larger samples such as tissue slices from the brain. However, the density of measurement spots limits the spatial resolution of current spatial barcoding methods, ranging from multicellular to subcellular. In addition, shotgun sequencing inevitably suffers from PCR amplification bias (61), as well as "dropout" when sequencing read depth is insufficient (62). Thus far, we have provided an overall picture of different spatial transcriptomic methods and their characteristics. Because this review focuses on analysis and visualization of spatial transcriptomics, readers who wish to understand the experimental details can refer to comprehensive reviews elsewhere (63).

**Table 2:** Current experimental methods for spatial transcriptomic profiling

| Method | Type | Resolution | Genes | Reference |
|---|---|---|---|---|
| **Visium** | Spatial barcoding | 55μm | Whole transcriptome | (47) |
| **Slide-seq** | Spatial barcoding | 10μm | Whole transcriptome | (48, 49) |
| **HDST** | Spatial barcoding | 2μm | Whole transcriptome | (50) |

| DBiT-Seq | Spatial barcoding | 10μm | Whole transcriptome | (51) |
|---|---|---|---|---|
| Seq-scope | Spatial barcoding | 0.6μm | Whole transcriptome | (52) |
| Stereo-seq | Spatial barcoding | 500nm | Whole transcriptome | (53) |
| seqFISH | *in situ* hybridization | single-molecule | >10,000 | (54, 55) |
| merFISH | *in situ* hybridization | single-molecule | $100 - 1,000$ | (56, 57) |
| STARmap | *in situ* sequencing | single-cell | $160 - 1020$ | (58) |
| FISSEQ | *in situ* sequencing | subcellular | $\sim 8000$ | (59) |

## 3 Preprocessing

Spatial transcriptomic datasets add a new dimension to transcriptomic analyses. Spatial coordinates of cells enable novel analyses such as spatial differential expression (34) and cell-cell interaction (41). Similar to single-cell RNA-seq datasets, a spatial transcriptomic dataset can be represented by a gene-by-cell count matrix. A second matrix of coordinates is attached to the cell dimension of the count matrix to represent spatial information. Comprehensive toolkits such as Space Ranger can process raw sequence reads into count matrices. Taking a microscope image and FASTQ files as input, Space Ranger can perform alignment, tissue and fiducial detection, barcode/UMI counting, and feature-spot matrices generation.

Various preprocessing steps may be performed prior to any analysis. First, genes and cells may be filtered based on a threshold specific to the dataset. For example, a cell may be removed if it has (i) less than 1,000 expressed genes or (ii) a high proportion of mitochondria RNA. A gene may be removed if it is detected in less than ten cells (45, 64). Transformation of count data may be performed according to downstream modeling assumptions. Methods modeling raw counts do not require any transformation (36). Otherwise, gene expression per cell may be normalized to have the same total library size such that expression levels are comparable across cells. The gene expression matrix may then be log-transformed and be regressed against confounders such as batch effect, percentage of mitochondria genes, and other technical variations. Although preprocessing steps mentioned above are widely adopted, the exact configuration should follow input data modality and modeling assumptions, and there is no one-size-fits-all strategy.

### 3.1 Gene expression normalization

Current spatial transcriptomic techniques introduce unwanted technical artifacts. Raw data commonly exhibit spot-to-spot variation and high dropout rates, which may impact downstream analyses. Several normalization strategies have been created to address these challenges. Due to the similarity between spatial transcriptomics and scRNA-seq, many normalization methods for spatial transcriptomics data are inspired by scRNA-seq studies.

A widely-used normalization tool is scran, a method based on the summation of expression values and deconvolution of pooled size factors (21). In the first step, expression values of all cells in the data set are averaged to serve as a reference. The cells are then partitioned into different pools, where the summation of expression values in each pool is normalized against the reference to generate a pool-based size factor. A linear system can be constructed by repeating the above normalization over multiple pools. Finally, the normalized cell-based counterparts can be calculated by solving the linear system with standard least-squares methods, i.e., deconvolving the pool-based size factor to individual cells. By representing the individual cells with multiple pools of cells, scran is capable of avoiding estimation inaccuracy in the presence of stochastic zeroes and is robust to differentially expressed genes. Similar to scran, a number of methods adopt the global scale factor strategy, where one normalization factor is applied to each cell, and all genes in this cell share the same factor. When the relationship between transcript-specific expression and sequencing depth is not shared across genes, such strategy will likely lead to overcorrection for weakly and moderately expressed genes. To address the problem, Bacher et al. proposed SCnorm, a quantile-regression based method that can estimate the dependence of expression on sequencing depth for each gene (22). Then genes are grouped based on the similarity of dependence, and a second quantile regression is used to estimate a shared scale factor within each group.

Lytal et al. conducted an empirical survey to evaluate the effectiveness of seven single-cell normalization methods. Based on the experimental results over several real and simulated data sets, the study concludes that there is no "one-size-fits-all" normalization technique for every data set (65). Further, Saiselet et al. investigated whether normalization is warranted for spatial transcriptomic datasets. They discovered that variation of total read counts is related to morphology and local cell density. Therefore, total counts per spot are biologically informative and do not necessarily need to be normalized out (66).

## 4    Analysis and visualization in the expression domain

A first step in the spatial transcriptomic analysis is to identify the cell type (for datasets of single-cell resolution) or cell mixture (for datasets of multicellular resolution) of each spatial unit or spot. Cell type identification usually starts with the dimensionality reduction technique to reduce time and space complexity for downstream analysis. The reduced representations are used to cluster cells based on the assumption that cells of the same type fall into the same cluster.

### 4.1    Clustering

The selection of clustering techniques is critical for obtaining good clustering results. Certain methods with assumptions about cluster shapes may not be suitable for spatial genomic data. For example, K-means clustering assumes that the shapes of clusters are spherical and that clusters are of similar size (67), and Gaussian mixture models assume that points with each cluster follow a Gaussian distribution (68). These assumptions are rarely satisfied by spatial transcriptomic data.

Agglomerative clustering methods are a class of methods that iteratively aggregate data points into clusters. These methods do not carry assumptions about the shape and size of clusters. At each iteration, data points are aggregated to optimize a pre-defined metric. Popular agglomerative clustering methods include hierarchical agglomerative clustering (69) and community detection methods such as Louvain

(23) and Leiden (24) algorithms. Hierarchical agglomerative clustering is initialized by treating each point as its own cluster. Each iteration aggregates two clusters with the closest distance to form a new cluster until no clusters can be merged. Community detection methods, i.e., Louvain (23) and Leiden (24) algorithms, have seen wide adoption in the single-cell and the spatial transcriptomics community. Both algorithms try to iteratively maximize the modularity, which can be understood as the difference between the number of observed and expected edges. Intuitively, a tightly connected community or cluster should have a large number of observed edges relative to the expected number of edges. The Louvain algorithm is initialized by assigning each node to its own community. At each iteration, each node moves from its own community to all neighboring communities, and changes in modularity are calculated. The node is moved to the community, which results in the largest increase in $H$. At the end of each iteration, a new network is built by aggregating all nodes within the same community, and a new iteration begins. The procedure will terminate when the increase in $H$ can no longer be achieved.

These general-purpose methods can be combined into more sophisticated pipelines tailored towards single-cell clustering. SC3 is an ensemble clustering method in which multiple clustering outcomes are merged into a consensus. SC3 first calculates distance matrices using the Euclidean distance, as well as Pearson and Spearman correlations. Spectral clustering is performed on these distance matrices with a varying number of eigenvectors. These results were combined to assign a consensus cluster membership to each point (25). Seurat uses a smart local moving (SLM) algorithm (43) to perform modularity-based clustering. Seurat first constructs a distance matrix based on canonical correlation vectors and a shared nearest neighbor (SNN) graph based on the distance matrix. The SNN graph is used as an input to the SLM algorithm to find clusters (70). SIMLR calculates a distance matrix as a weighted sum of multiple distance kernels and solves for a similarity matrix to minimize the product between the distance and similarity matrices. To ensure a fixed number of connected components, SIMLR uses constrained optimization to encourage a block diagonal structure in the similarity matrix (26).

## 4.2 Identification of cell types

Identification of cell types starts by defining cell-type specific genes or marker genes. A straightforward approach is to perform differential expression analysis (71, 72) between all pairs of clusters. Genes that are consistently over-expressed in one cluster are considered the cluster's marker genes. This is the approach implemented in scran (64) and Mast (73).

Another method, scGeneFit, uses a label-aware compression method to find marker genes (27). Given cell-by-gene expression matrix and corresponding cell labels inferred from clustering results, scGeneFit finds a projection onto a lower-dimensional space, in which cells with the same labels are closer in the lower-dimensional space than cells with different labels. The projection is constrained such that the axes in the lower-dimensional space align with a single gene. Therefore, the marker genes will be the set of axes in the lower-dimensional space that best conserves label structures. The marker genes can then be matched with an expert-curated list of cell-type specific genes to infer cell types (74, 75). Other methods directly map unknown cell types onto a reference dataset, bypassing the target gene identification step. Scmap projects the query cells onto the reference cell types from other experiments and datasets (28). The known reference cluster is represented by its centroid, and the projection is carried out by a fast approximate k-nearest-neighbor (KNN) search by cluster using product quantization (76), where a similarity matrix between the query cell and reference clusters is used as the distance in KNN search. Another reference-based method is SingleR (29). The method proceeds by first identifying variable genes among cell types in the reference set. Next, SingleR calculates the Spearman correlation between each single cell and the reference variable genes. Multiple correlation coefficients within each cell type are aggregated to form one correlation per reference cell type per single cell. Only the top 80% of correlation values are selected to remove random noise. In the fine-tuning step, the correlation analysis is iteratively re-run but only for the top cell types from the previous step, and the lowly correlated cell types are removed. Eventually, the cell type with the top correlation is assigned to the query single-cell. Using SingleR, the authors identified a novel disease-associated macrophage subgroup between monocyte-derived and alveolar macrophages. Cortal et al. proposed a clustering-free multivariate statistical method named Cell-ID for gene signature extraction and cell identification (30). Cell-ID first performs a dimensionality reduction on the cell-by-gene expression matrix using the multiple correspondence analysis (MCA). Both cells and genes are simultaneously projected in a common low-dimensional space, where the distance between a gene and a cell represents the specific degree between them. According to the distance, Cell-ID can build up a gene-rank for each cell, and the top-ranked genes are defined as the cell's gene signature. With the gene signature of the query cell, Cell-ID can perform automatic cell type and functional annotation via the hypergeometric tests against reference marker gene lists and/or gene signatures of reference single-cell datasets. The authors demonstrated the consistently reproducible gene signatures across diverse benchmarks, which helps to improve biological interpretation at the individual cell level. Unlike the above approaches, JSTA uses deep learning for cell-type identification and incorporates three distinct and interactive components: a segmentation map and two deep neural network-based cell type classifiers for pixel-level and cell-level classification (31). JSTA first trains a taxonomy-based cell-level classifier with the external data from the Neocortical Cell Type Taxonomy (NCTT) set (77). Then the segmentation map and pixel-level classifier are iteratively refined with an expectation-maximization (EM) algorithm. Specifically, the segmentation map is initialized by a classical image segmentation algorithm watershed (78) and paired with the trained cell-level type classifier to predict the current cell (sub)types. Given the local mRNA density at each pixel as the input, the pixel-level classifier is optimized to closely match each pixel's current cell type assignment. Next, the updated pixel-level classifier reclassifies the cell types of all border pixels, and the resulting segmentation map requires an update of the cell-level classification, which further triggers an update of pixel-level classifier training. This learning process is repeated until convergence. The eventual segmentation map tends to maximize consistency between local RNA density and cell-type expression priors. Abdellaal et al. benchmarked 22 broadly used cell identification methods on 27 publicly available single-cell RNA data. Interested readers are referred to (79).

### 4.3   Visualization of gene expression in low dimensions

The identified clusters can be visualized to ensure cells assigned to the same cluster are close in expression space. Dimensionality reduction techniques are necessary to project the high dimensional data into 2D or 3D. Principal component analysis (PCA) is widely adopted in the single-cell and spatial transcriptomic literature (80). This method identifies linear combinations of the original dimensions, or principal components (PC), that maximize the projection variance from data points onto the principal components (Figure 2A). The principal components can be computed in an iterative way: the first PC can point in any direction to maximize the variance of projections, and each subsequent PC is orthogonal to previous PCs (81).

In contrast to PCA, manifold learning is a class of non-linear dimensionality reduction techniques that aims to project the data to a lower dimension while maintaining the distance relations in the original high-dimension space; points close to each other in the original space will be close in the low-dimensional space (Figure 2B). Uniform manifold approximate and projection (UMAP) and t-distributed stochastic neighbor embedding (t-SNE) are two manifold learning methods widely adopted in single-cell and spatial transcriptomic literature (32, 33). Both methods follow a two-step procedure. In the first step, a similarity matrix is computed based on a pre-defined distance metric. In the second step, all data points are placed in a low-dimensional Euclidean space such that the structure of the similarity matrix is preserved. This step is initialized by randomly placing data points in the low-dimensional space. At each iteration, data points are moved according to the similarity matrix from the high-dimensional space; points with high similarity in the high-dimensional space will attract, and those with low similarity will repel. Because optimization is done iteratively, UMAP and t-SNE results are stochastic and vary between runs. Random seeds are needed for reproducibility. The two methods differ in their construction of similarity matrix. In t-SNE, a distance matrix is calculated according to probability density functions (PDF) of the Gaussian distribution in the high-dimension space and PDFs of the t-distribution in the low-dimension embedding. In UMAP, an adjacency matrix is constructed by extending a sphere whose radius depends on the local density of nearby points; two points are connected if their spheres overlap. In practice, UMAP is faster than t-SNE and tends to preserve the high-dimensional structure better.

## 5    Analysis and visualization in the spatial domain

An important question in spatial transcriptomic data analysis is to identify genes whose expression follow coherent spatial patterns. Genes with spatial expression patterns are critical determinants of polarity and anatomical structures. For example, the gene *wingless* is a member of the *wnt* family that plays a central role in anterior-posterior pattern generation during the embryonic development of *Drosophila melanogaster*. It is expressed in alternating stripes across the entire embryo (82). Another example is the neocortex of mammalian brains, which contain six distinct layers. Each layer consists of different types of neurons and glial cells that express cell-type specific marker genes (83). Spatial transcriptomic data enables unbiased transcriptome-wide identification of spatially expressed genes, but it is excessively labor-intensive to visually examine all genes. This prompted the development methods including SpatialDE (34), trendsceek (35), and Spark (36).

### 5.1    Identifying genes with spatial expression patterns

SpatialDE (34) uses a Gaussian process to model gene expression levels. Intuitively, a Gaussian process model treats all data points as observations from a random variable that follows a multivariate Gaussian (MVN) distribution (84). To test whether expression levels follow a spatial pattern, the authors specify a null model, in which the covariance matrix is diagonal, and an alternative model, in which the covariance matrix follows a radial basis function kernel:

$$K\left(x_i, x_j\right) = \exp\left(-\gamma \left\| x_i - x_j \right\|^2\right)$$

<div align="right">**Eq. 1**</div>

where $K(x_i, x_j)$ is the covariance between i-th and j-th measurement; $x_i$ and $x_j$ represent the spatial coordinates of the i-th and the j-th measurement; $\gamma$ is a scale factor. Intuitively, the Gaussian kernel describes a spatial relationship in which nearby points have similar expression values. This kernel assumes that cells of similar origins tend to neighbor each other in space. A likelihood ratio test can be done by comparing the likelihood of the null and the alternative model. Because SpatialDE is a Gaussian process model, the expression values must be log-transformed which decreases power.

Trendsceek (35) uses a marked point process model in which each point of measurement, or a spot, is treated as a point process, and each point is marked with a gene expression value. To decide whether a gene whose expression follows a spatial pattern, trendsceek test whether the probability of finding two marks given the distance between two points deviates from what would be expected if the marks were randomly distributed over points. To calculate the null distribution given no spatial pattern, trendsceek implements a sampling procedure in which marks are permuted with the location of points fixed. In practice, such sampling procedure is computationally expensive and makes trendsceek only suitable for small datasets.

Spark (36) uses a generalized linear spatial model (GLSM) to directly model count data (85, 86), which results in better power than SpatialDE. A simplified model is presented below:

$$y(s) \sim Poisson\big(\lambda(s)\big)$$ 
**Eq. 2**

$$\log\big(\lambda(s)\big) = x(s)^T \beta + b(s) + \epsilon$$ 
**Eq. 3**

$$b(s) \sim MVN\big(0, \tau K(s)\big)$$ 
**Eq. 4**

Where $y(s)$ is the gene expression of sample $s$. $\lambda$ is a Poisson rate parameter, which is modeled as a linear combination of three terms. The first term $x(s)$ represents covariates such as batch effect and library size for sample $s$. The second term $b(s)$ is the spatial correlation pattern modeled as a Gaussian process. The last term $\epsilon$ is random noise. To determine whether a gene follows a spatial pattern, Spark tests whether $\tau = 0$. Parameter estimation is difficult due to the random effects. *Monte Carlo* methods are the gold standard for parameter estimation for GLSM but are computationally expensive. Instead, the authors developed a penalized quasi-likelihood (PQL) estimation procedure to make computation tractable for large datasets (86-88). Spark produces well-calibrated p-values and is more powerful than trendsceek and SpatialDE through a series of simulation experiments.

## 5.2 Identification of spatial domains

Spatially coherent domains often underly important anatomical regions (Figure 3A). A motivating example is the histological staining of cancer tissue slides. Cancer regions and normal tissues can be visually distinguished due to differential affinities to staining agents. This enables pathologists to grade and stage individual cancer tissue slides according to the location and size of the cancer regions (89). Spatial transcriptomics enables histology-like identification of spatial domains. Regular histology slides can be visualized conveniently with RGB pixels. In contrast, spatial transcriptomic data cannot be directly visualized because each spot (i.e., pixel) in spatial transcriptomic data has a dimension equal to the number of genes. This prompts the development of methods to detect spatial domains, including BayesSpace (40), SpatialDE (34), and a hidden Markov random field (HMRF) method (37).

The three methods share a common assumption that hidden spatial domains can be described by latent variables, which are not directly observed but can be inferred from observed gene expression values. However, these methods use different modeling assumptions to infer latent variables. Zhu et al. (37) developed an HMRF-based method, a widely adopted model in the image processing community to identify patterns in 2D images (90, 91), to identify spatial domains. An HMRF has two components: it uses a Markov random field to describe the joint distribution of latent variables and a set of observed examples that depends on them. The latent variables are assumed to satisfy the Markov property, in which any node in the network is conditionally independent of other nodes given its neighbors. Following this assumption, a Markov random field of latent variables can be decomposed into a set of subgraphs, called cliques, which gives rise to the observed gene expression. The parameters of the model by Zhu et al. are estimated with an EM algorithm (92, 93).

Both SpatialDE (34) and BayesSpace (40) model observed gene expression values as a mixture of Gaussian random variables. The means of the Gaussian random variables are determined by the spatial domain membership. In SpatialDE, the mean expression value of each spatial domain is described by a Gaussian process, whose covariance follows a radial basis function kernel. The observed expression follows a Gaussian distribution centered around the mean expression value of a given spatial domain. The posterior distribution of parameters and the latent spatial domain membership is estimated by variational inference. Different from SpatialDE, BayesSpace uses a diagonal matrix to model the covariance of the mean expression of each spatial domain. The observed gene expression is modeled as a Gaussian random variable centered around the mean expression and has a diagonal covariance matrix modeled as a Wishart random variable. BayesSpace uses a Markov chain Monte Carlo (MCMC) method to estimate model parameters (94).

While the above methods consider spatial genes and spatial domain as two separate tasks, SpaGCN proposed a graph convolutional network-based (GCN) approach to address these two tasks jointly (38). With the integration of gene expression, spatial location, and histology information, SpaGCN models spatial dependency of gene expression for clustering analysis of spatial domains and identification of domain enriched spatial variable genes (SVG) or meta genes. SpaGCN first converts the spatial transcriptomics data into an undirected weighted graph of spots, and the graph structure represents the spatial dependency of the data. Next, a GCN (95) is utilized to aggregate gene expression information from the neighboring spots and update every spot's representation. Then, SpaGCN adopts an unsupervised clustering algorithm (96) to cluster the spots iteratively, and each identified cluster will be considered as a spatial domain. The resulting domains guide the differential expression analysis to detect the SVG or meta genes with enriched expression patterns in the identified domains.

### 5.3 Spot deconvolution and super-resolution

Because spots in the spatial transcriptomic dataset may not correspond to cell boundaries, several additional features can be included when plotting on the spatial domain. When the spatial transcriptomic measurement technology has a multicellular resolution, spots can be decomposed into constituent cell types. A 2D array of pie charts can be used to represent the cell types' percentages of spots, as demonstrated in DSTG (Figure 3B). To enable the investigation of cellular architecture at higher resolution, DSTG uses a GCN to uncover the cellular compositions within each spot (39). DSTG first leverages single-cell RNA-seq data to construct pseudo spatial transcriptomic (pseudo-ST) data by selecting two to eight single cells from the same tissue and combining their transcriptomic profiles. This pseudo-ST data is designed to mimic the cell mixture in the real spatial transcriptomic data and provide the basis for model training. Via canonical correlation analysis, DSTG identifies a link graph of spots with the integration of the pseudo-ST data and the real spatial transcriptomic data. A GCN (95) iteratively updates the representation of each spot by aggregating its neighborhoods' information. The GCN model is trained in a semi-supervised manner, where the known cell compositions of the pseudo-ST nodes are served as the labeled data, and the real spatial transcriptomic nodes are the prediction targets. The resulting cell type proportions can be displayed as a pie chart at each spot (Figure 3B).

While cell type deconvolution provides an estimation of cellular constituents, it does not directly increase the resolution of the dataset. BayesSpace uses a Bayesian model to increase the resolution to the subspot level, which approaches single-cell resolution with the Visium platform (Figure 3C). The model specification is similar to the spatial domain detection model described above, except that unit of analysis is the subspot rather than the spot. Since gene expression is not observed at the subspot level, BayesSpace models it as another latent variable and estimates it using MCMC. The increase in resolution is different across measurement technology. For square spots (47), BayesSpace by default divides each spot into nine subspots. For hexagonal spots like Visium, they are divided into six subspots by default. The subspots can be visualized in Euclidean space similar to regular spots.

## 5.4   Visualization in Euclidean space

After obtaining spatial genes and domains, visualization in the Euclidean space is relatively straightforward. Spatial genes can be visualized by plotting their log-transformed expression values. Spatial domains can be colored by mean expression values or by their identities. Several packages such as Giotto (44), Scanpy (45), Seurat (97), and Squidpy (46) provide functionalities to plot spatial transcriptomic data in Euclidean space.

## 6   Analysis and visualization in the interaction domain

Cell signaling describes the process in which cells send, receive, process, and transmit signals within the environment and with themselves. Based on the signaling distance and the sender-receiver identities, cell signaling can be classified into autocrine, paracrine, endocrine, intracrine, and juxtacrine (98). It serves critical functions in development (99), immunity (100), and homeostasis (101) across all organisms. For example, the Hedgehog signaling pathway is involved in tissue patterning and orientation, and aberrant activations of hedgehog signaling lead to several types of cancers (102). Single-cell datasets enable correlation analysis to unravel cell-to-cell interaction (103-105). Due to the lack of spatial information, single-cell analysis cannot distinguish short-distance (juxtacrine and paracrine) and long-distance (endocrine) signaling. Spatial transcriptomic datasets provide the spatial coordinate of each cell or spot and enable spatial dissection of cell signaling.

## 6.1   Cell-to-cell interaction

Cell signaling frequently occurs between cells in spatial proximity. Giotto takes spatial proximity into consideration to identify cell-to-cell interaction. It first constructs a spatial neighborhood network to identify cell types that occur in spatial proximity. Each node of the network represents a cell, and pair of neighboring cells are connected through an edge. The neighbors of each cell can be determined by extending a circle of a predefined radius, selecting the k-nearest neighbors, or constructing a Delaunay network (106). Cell types connected in the network more than expected are considered interacting. Giotto permutes the cell type labels without changing the topology of the network and calculates the expected frequencies between every pair of cell types. P-values are derived based on where the observed frequency falls on the distribution of expected frequencies.

Another method, SpaOTsc, leverages both single-cell and spatial transcriptomic data for a comprehensive profile of spatial interaction (41). It uses an optimal transport algorithm to map single-cell to spatial transcriptomic data. An optimal transport is a function that maps a source distribution to a target distribution while minimizing the amount of effort with respect to a predefined cost function (107). SpaOTsc generates a cost function based on the expression profile dissimilarity of shared genes across the single-cell and the spatial transcriptomic datasets. The optimal transport plan maps single cells onto spatial locations. SpaOTsc then formulates cell-to-cell communication as a second optimal transport problem between sender and receiver cells. The expression of ligand and receptor genes are used to estimate sender and receiver cells, and the spatial distance is the cost function. The resultant optimal transport plan represents the likelihood of cell-to-cell communication.

## 6.2    Ligand-receptor pairing

Another aspect of cell signaling is the pairings between ligands and receptors. Giotto identifies ligand-receptor pairs whose mean expression is higher than expected. To obtain the observed expression of ligand-receptor pairs for a pair of cell types, Giotto averages the expression of ligand in all sender cells and the expression of receptors in all receiver cells in proximity of the sender cells. Giotto then permutes the location of cells to obtain an expected expression of the ligand-receptor pair. A p-value can be obtained by mapping the observed expression onto the distribution of expected expression. Different from Giotto, SpaOTsc uses a partial information decomposition (PID) approach to determine gene-to-gene interaction. Intuitively, PID decomposes the mutual information between multiple source variables and a target variable into unique information contributed by each source variable, redundant information shared by many source variables, and synergistic information contributed by the combination of source variables (108). SpaOTsc estimates the unique information from a source gene to a target gene that is within a predefined spatial distance, taking into consideration all other genes. Yuan *et al.* proposed a method called GCNG (42) to infer the extracellular gene relationship using Graph convolutional neural networks (GCN). Single-cell spatial expression data is represented as a graph of cells. Cell locations are encoded as a binary cell adjacency matrix with a selected distance threshold, and expression of gene pairs within each cell is encoded as corresponding node features. A GCN is used to combine the graph structure and node information as input and predict whether the studied gene pair can interact. The deep learning model is trained in a supervised manner, where positive samples are built from known ligand-receptor pairs, and negative samples are randomly selected from non-interacting genes. In addition to the methods specifically designed to utilize the spatial expression information for cell-to-cell interaction, many other tools developed for expression data without spatial information can also be applied to the spatial transcriptomic data. Interested readers can refer to a recent review of these methods (109).

## 6.3    Visualization of interactions between cells and genes

Cell-to-cell and gene-to-gene interactions are naturally represented as networks and correlation matrices (Figure 4). Integrative packages such as Giotto (44) provide functions to visualize cell-to-cell and gene-to-gene interactions as heatmaps, dot plots, or networks. A heatmap is a visual depiction of a matrix whose values are represented as colored boxes on a grid. With heatmaps, large blocks of highly connected cells or genes can be visually identified. A dot plot is similar to a heatmap, except that the boxes are replaced by dots of varying sizes. A dot plot can use both the size and the color of each dot to represent values in each interaction. Different from heatmaps and dot plots, networks use nodes to represent cells or genes and edges to represent their interactions. The widths and colors of edges can be used to describe the strength of interactions. Besides Giotto, the igraph package is widely adopted for network visualization and provides programming interfaces in R, python, C/C++, and Mathematica (110). Cytoscape is another widely used package to visualize complex network interaction. Its graphical user interface makes it easy to manipulate and examine nodes and edges in the network (111).

## 7    Discussions

Spatial transcriptomic technologies have made tremendous progress in recent years. Although earlier technologies are restricted by the number of profiled genes (9, 56, 57) or the spatial resolution (47), current methods can profile the whole transcriptome at single-cell or subcellular resolution (51-53). While available commercialized methods (Visium) cannot achieve cellular resolution, we believe newer technologies will soon be production-ready. As commercial platforms become more affordable, we believe the speed at which spatial transcriptomic datasets become publicly available will only accelerate. For example, phase two of the Brain Initiative Cell Census Network (BICCN) will map the spatial organization of more than 300,000 cells in the mouse's primary motor cortex (112). Large-scale projects to comprehensively profile spatial gene expression are currently limited, but we envision that these projects will expand in three directions. First, more model organisms will be profiled, enabling comparative analysis of cell types and their spatial organizations across evolution. Second, more organ and tissue types will be profiled for a comprehensive understanding of spatial expression architecture. Third, cell states (e.g. stimulated vs. resting) and disease states (cancer vs. normal) will be profiled to understand cellular activation and disease pathology.

As spatial transcriptomic datasets become more abundant, meta-analysis across published datasets will become commonplace. Methods to remove batch effects are needed to account for technical confounders across datasets. Unlike bulk and single-cell sequencing, batch effects in spatial transcriptomic data must account for correlation across space. Further, the batch effect may also occur on companion histology images, and methods to jointly analyze histology image and spatial transcriptomic data are required. Although several methods have been developed for batch effect removal in bulk (113, 114) and single-cell (115, 116) sequencing, it is still an under-explored area for spatial transcriptomics.

Histopathology is widely adopted across various domains of medicine and is considered the gold standard for certain diagnoses such as cancer staging (117). However, histology is limited by the type and number of cellular features delineated by staining agents. Spatial transcriptomics extends histology to test for both imaging and molecular features and may enable testing for oncogenic driver mutations critical for determining cancer subtypes. A recent method named SpaCell integrates both histology and spatial transcriptomic information to predict cancer staging (118). In this method, histological images are tiled into patches, where each patch corresponds to a spatial transcriptomic spot in a tissue. A convolutional neural network is used to extract image features from each patch, and combine the features with the spot gene count. A subsequent deep network is applied to predict the disease stages. We envision that spatial transcriptomics will become a diagnostic routine as it becomes more affordable and the clinical interpretation becomes more streamlined.

In this review, we surveyed state-of-the-art methods for spatial transcriptomic data analysis and visualization, and categorized them into three main categories according to the way their output is visualized. It is unlikely that we covered all available methods for spatial transcriptomics, but we hope this review will serve as a stepping stone and attract more researchers to this field.

## 8    Figure Captions

**Figure 1** An overview of spatial transcriptomic tasks. **(A)** Spatial transcriptomic datasets map gene expression measurements to their respective locations. **(B)** A spatial transcriptomic dataset can be analyzed in gene expression space, irrespective of spatial locations. Tasks such as clustering and cell-type identification fall into this category. **(C)** Spatial information can be used jointly with gene expression to detect spatial expression patterns and spatial domains. **(D)** These two sources of information can also be used to detect cell-cell and gene-gene interactions.

**Figure 2** Comparison between principal component analysis and t-SNE. **(A)** Principal component analysis iteratively identifies vectors that minimize the sum of squared distances to the direction of the vector. Each vector is orthogonal to all previously selected vectors. **(B)** t-SNE calculates a pairwise similarity based on the probability density function of the Gaussian distribution in the original high dimension space. The points are randomly projected to a low dimensional space and iteratively refined

so that the similarity in low dimension matches that of high dimension. At each iteration, similar pairs attract, and dissimilar pairs repel each other.

**Figure 3** Visualization of gene expression in the Euclidean space. **(A)** Spatially coherent genes and spatial domains can be visualized as 2D images. **(B)** Spot deconvolution methods estimate the proportion of each cell type within each spot. Pie charts are routinely used to represent cell type proportions within each spot. **(C)** Spot super-resolution methods estimate the cell type of sub-pixels based on correlation with neighbor spots. In this case, each spot of the original dataset is divided into nine spots in the super-resolved dataset.

**Figure 4** Identification of cell-cell and receptor-ligand interaction. **(A)** Cell-to-cell interaction can be identified by the correlation of gene expression values between cell pairs. **(B)** Receptor-ligand interaction can be identified by the correlation between genes in interacting cell types.

## 9    Conflict of Interest

We declare no conflict of interest.

## 10    Author Contributions

B.L. conceived the project. B.L. and Y.L. performed literature review. B.L. and Y.L. wrote the paper. Z.L. made the illustrations.

## 11    Funding

We have no funding source to disclose.

**A** Tissue sample

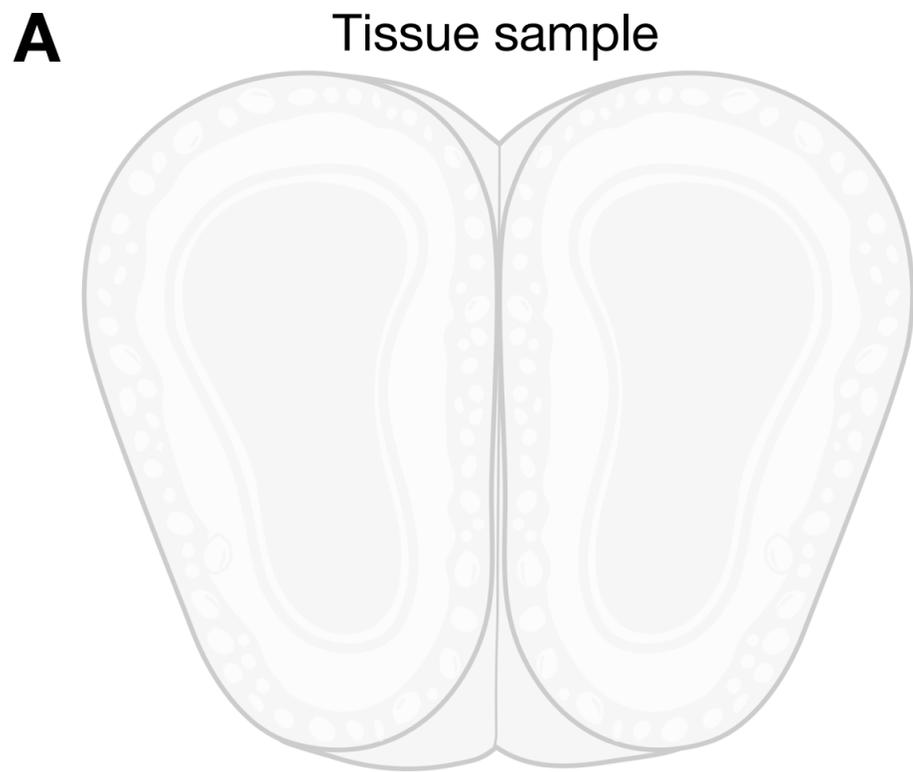

| x coordinates |
| y coordinates |
| (z coordinates) |

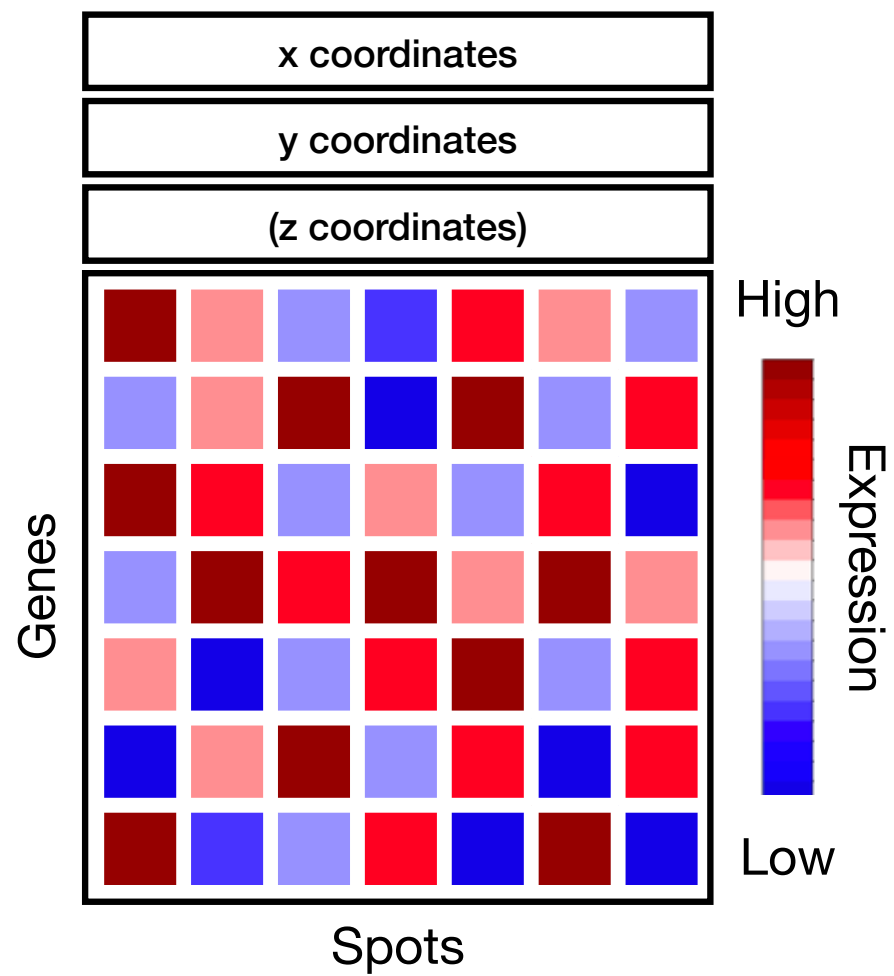

Genes

Spots

High

Expression

Low

**B** Clusters of spots

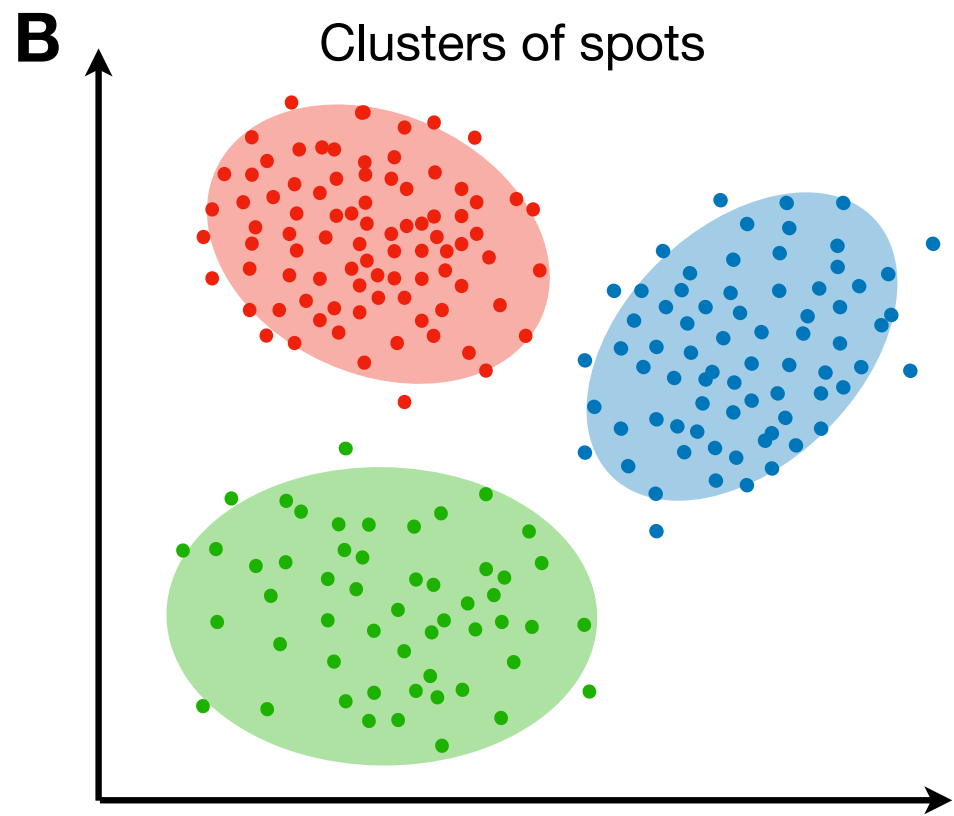

**C** Spatial domains

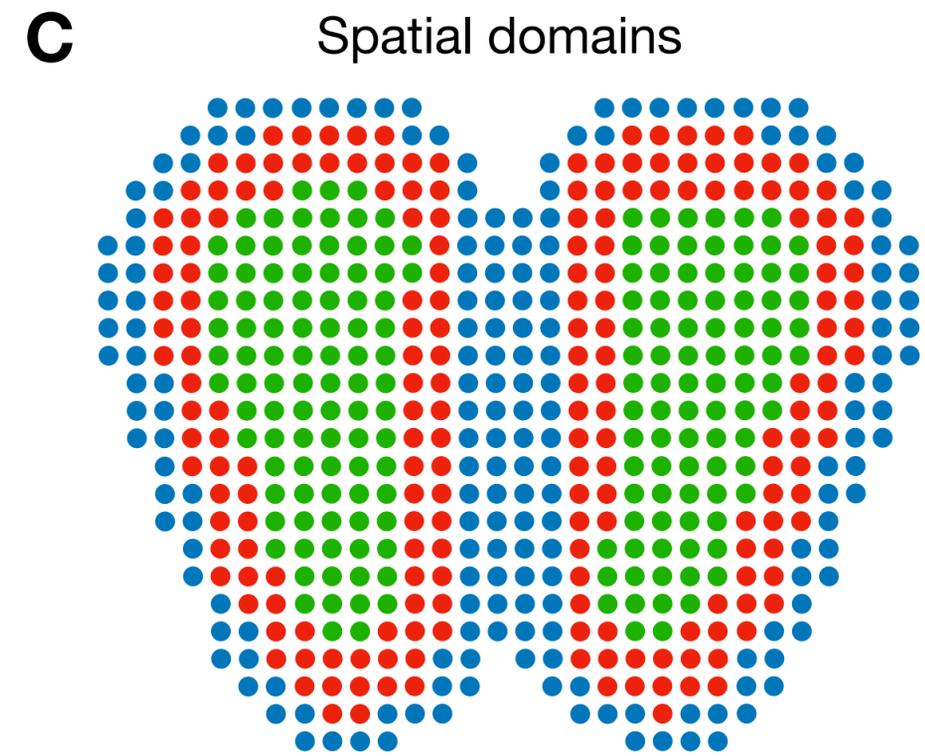

**D** Interaction network

Cell-cell communication

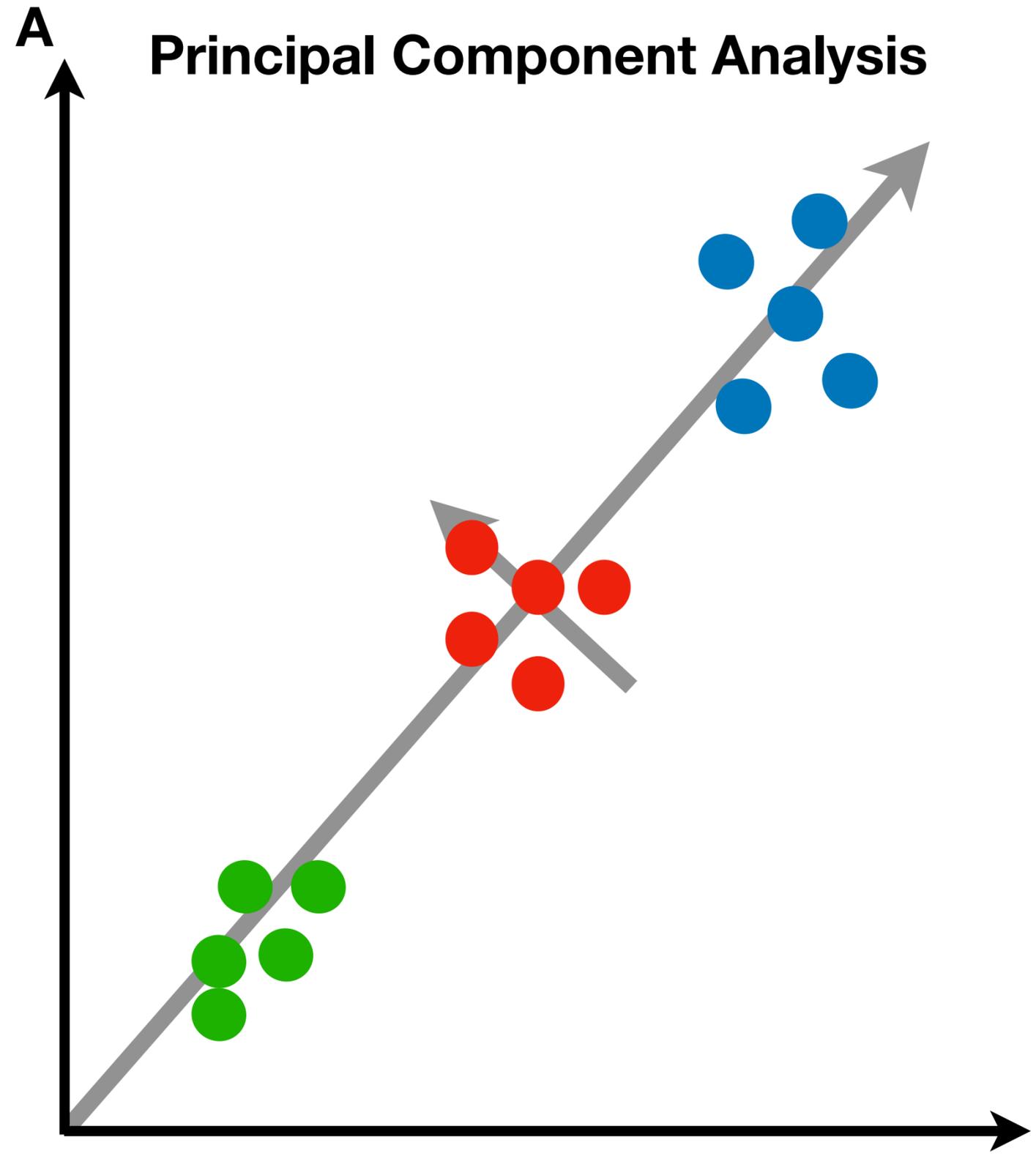

## A

**Principal Component Analysis**

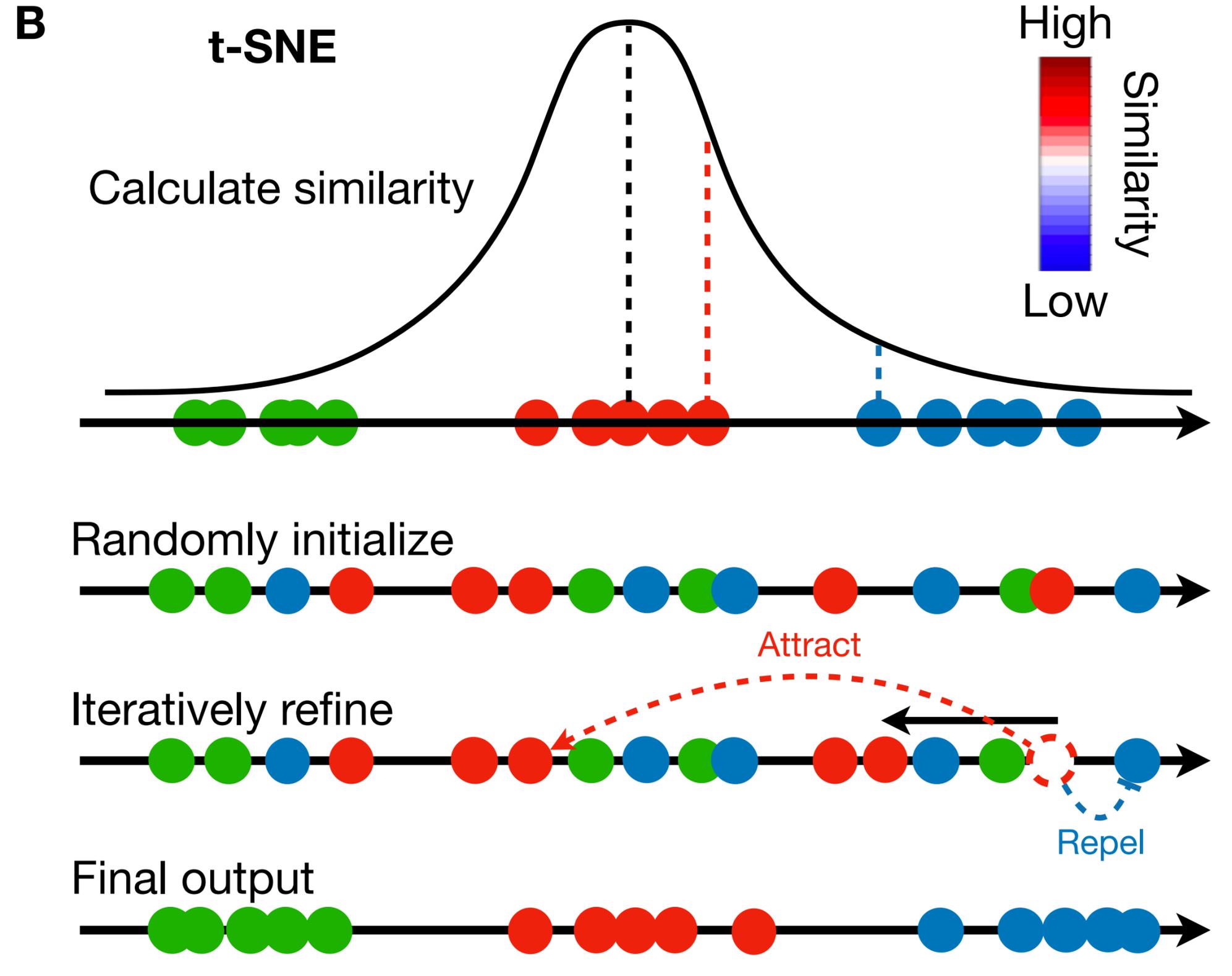

## B

**t-SNE**

Calculate similarity

High

Similarity

Low

Randomly initialize

Iteratively refine

Attract

Repel

Final output

**A** Spatial gene and domains

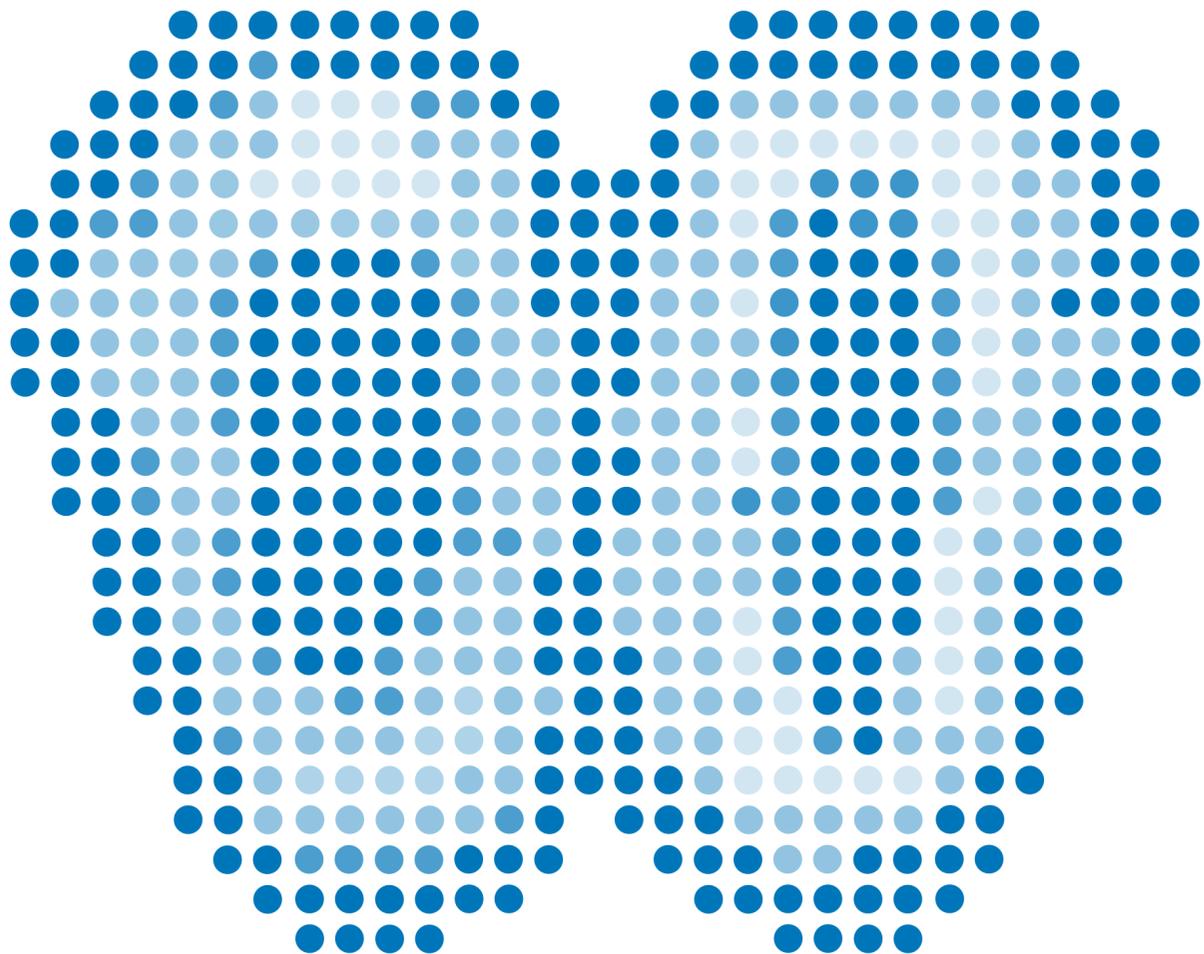

**B** Spot deconvolution

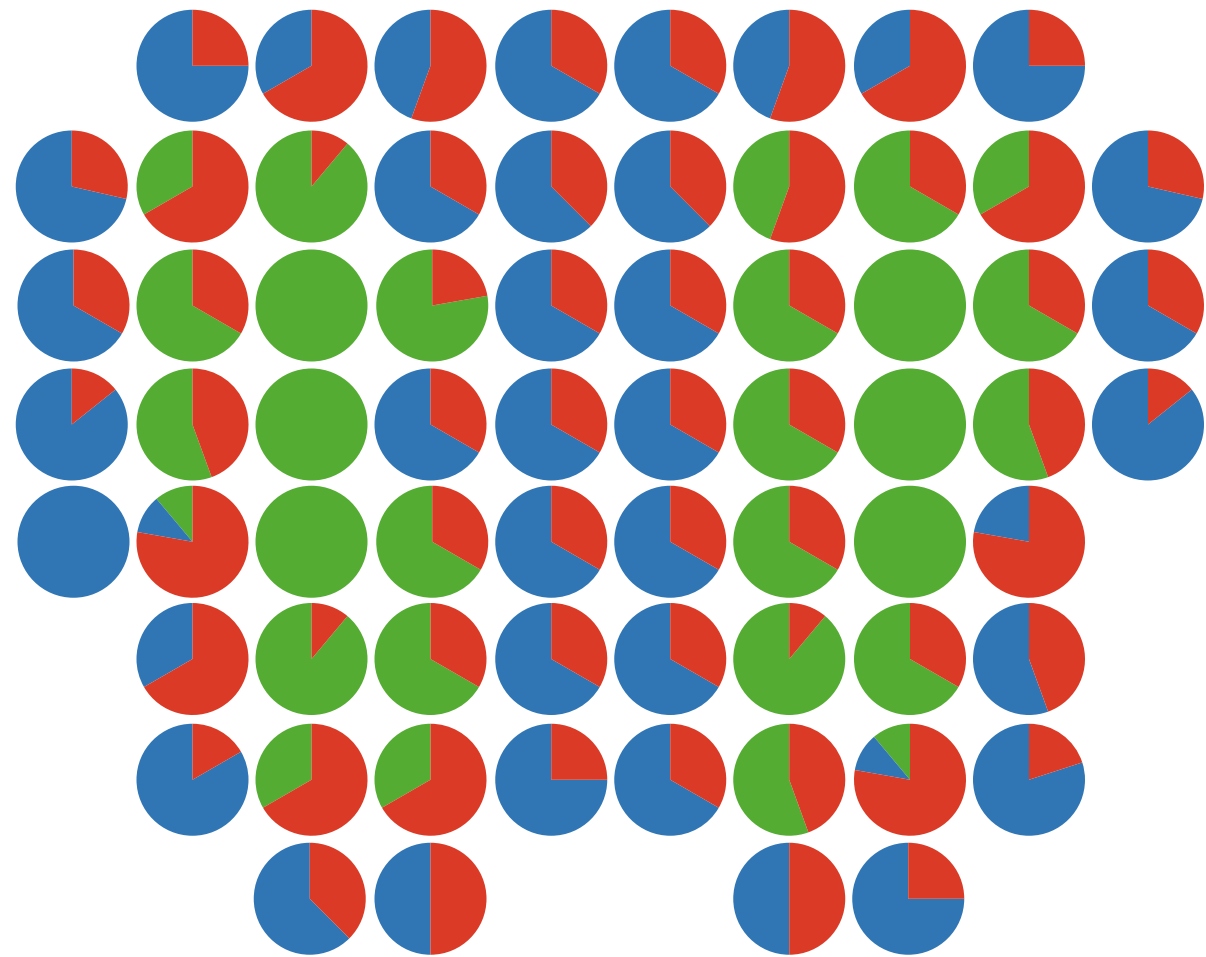

**C** Original dataset

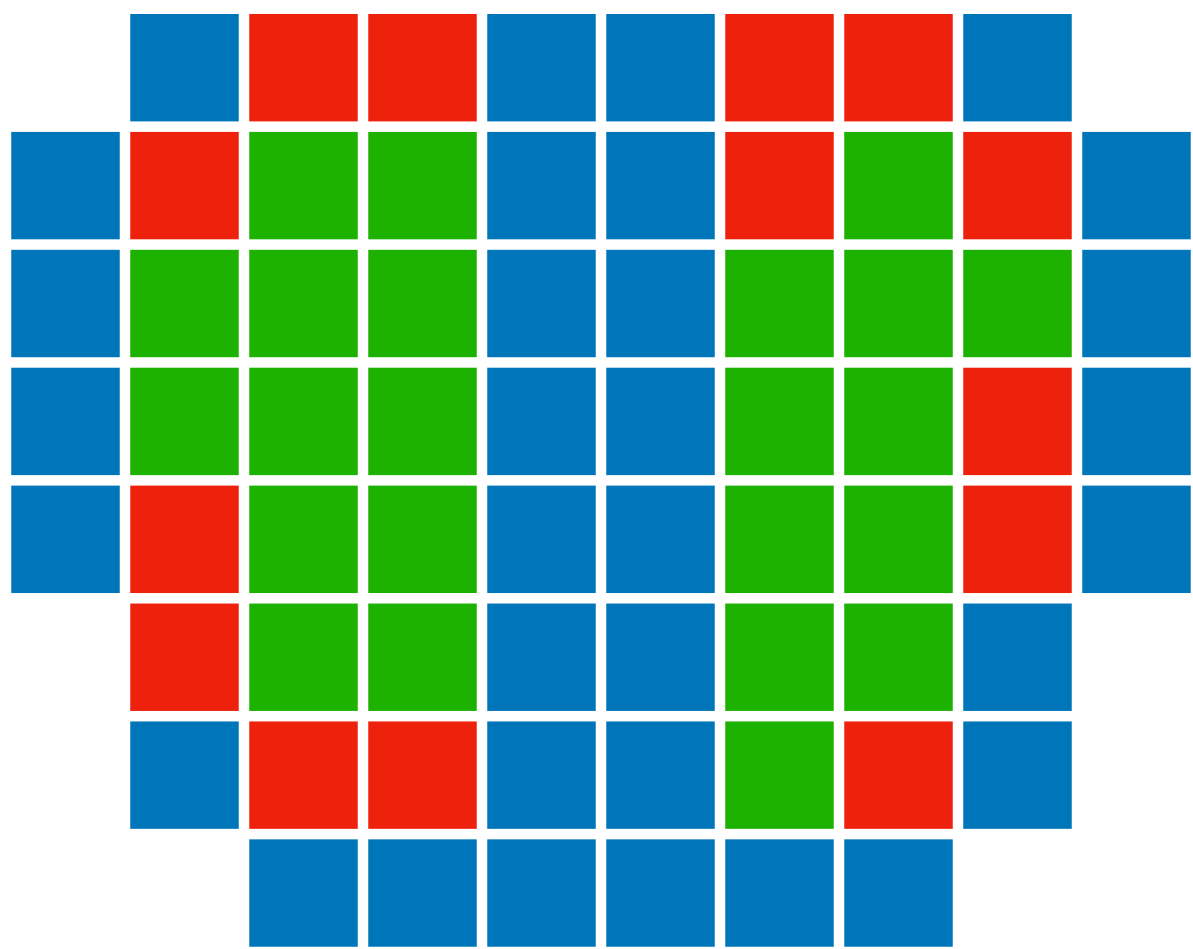

Super-resolved dataset

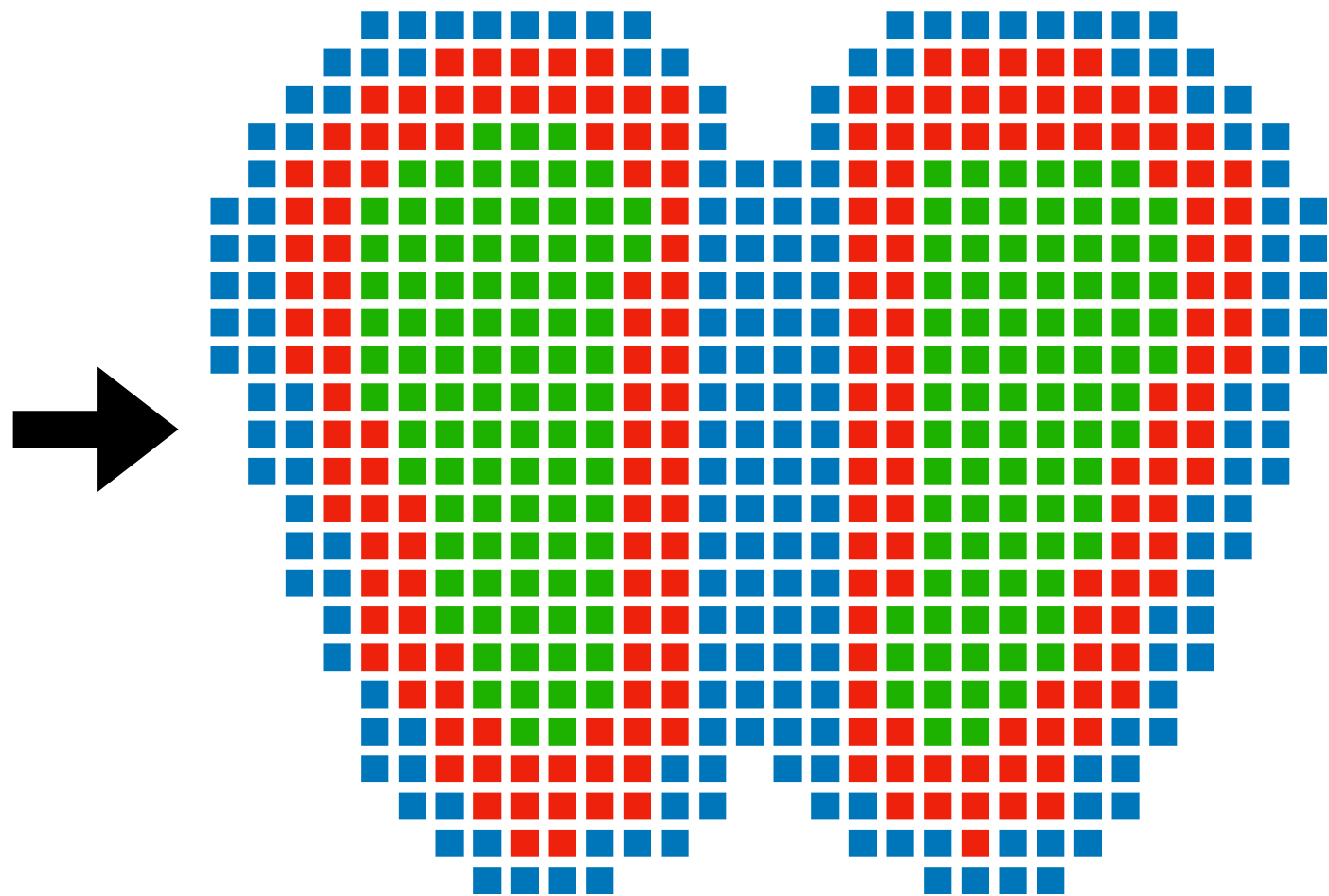

**A**

| x coordinates |
| y coordinates |
| (z coordinates) |

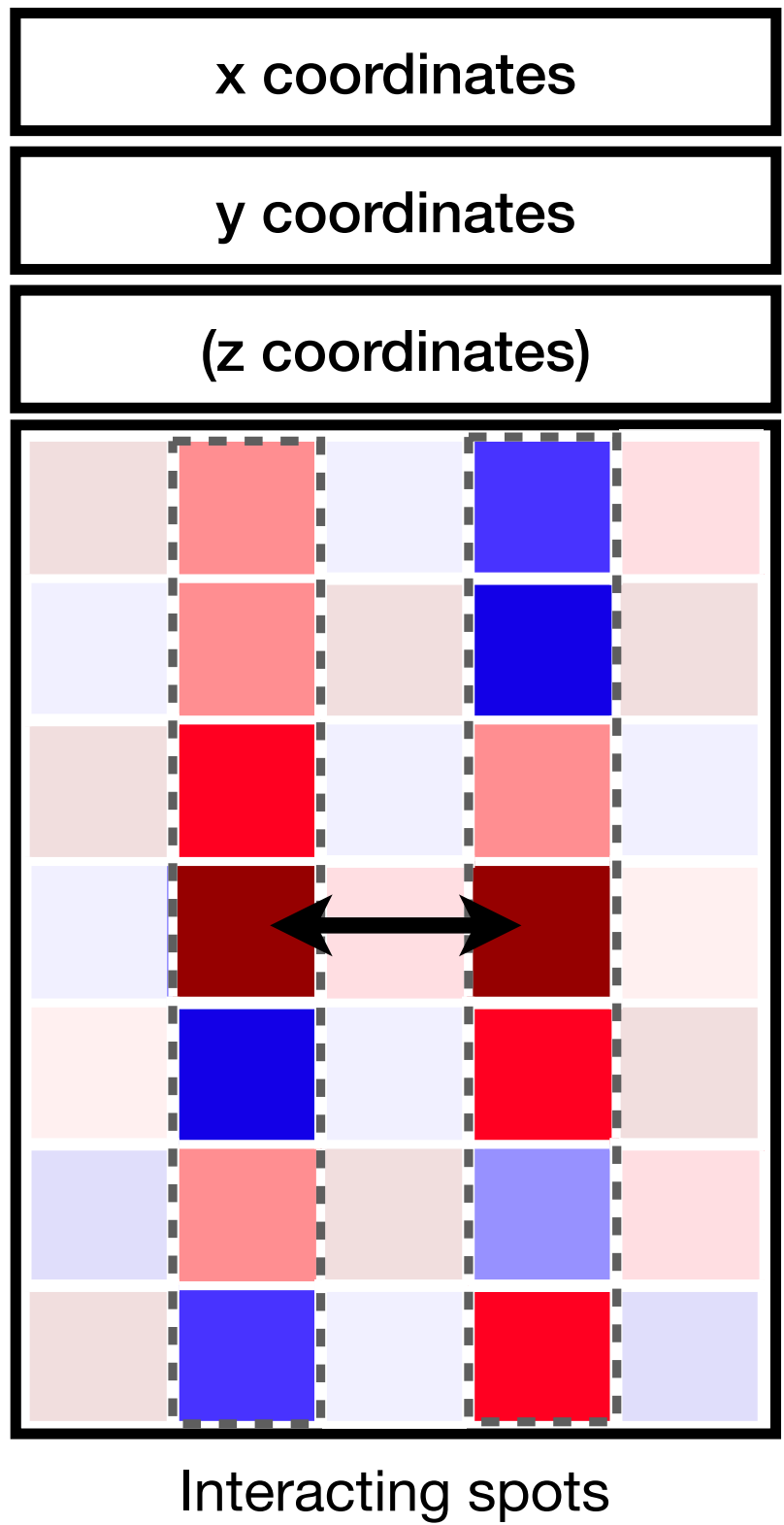

Interacting spots

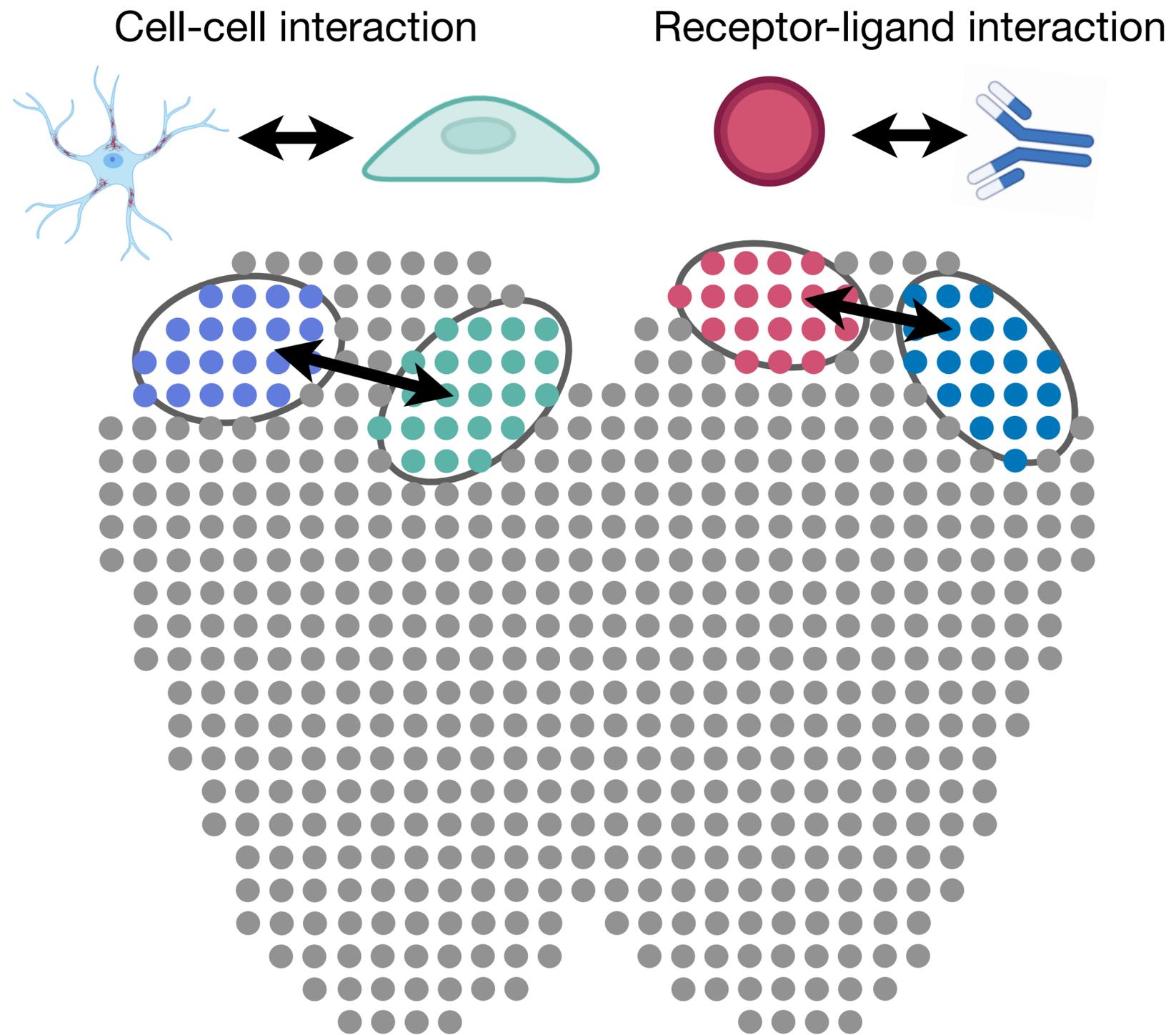

Cell-cell interaction

Receptor-ligand interaction

**B**

| x coordinates |
| y coordinates |
| (z coordinates) |

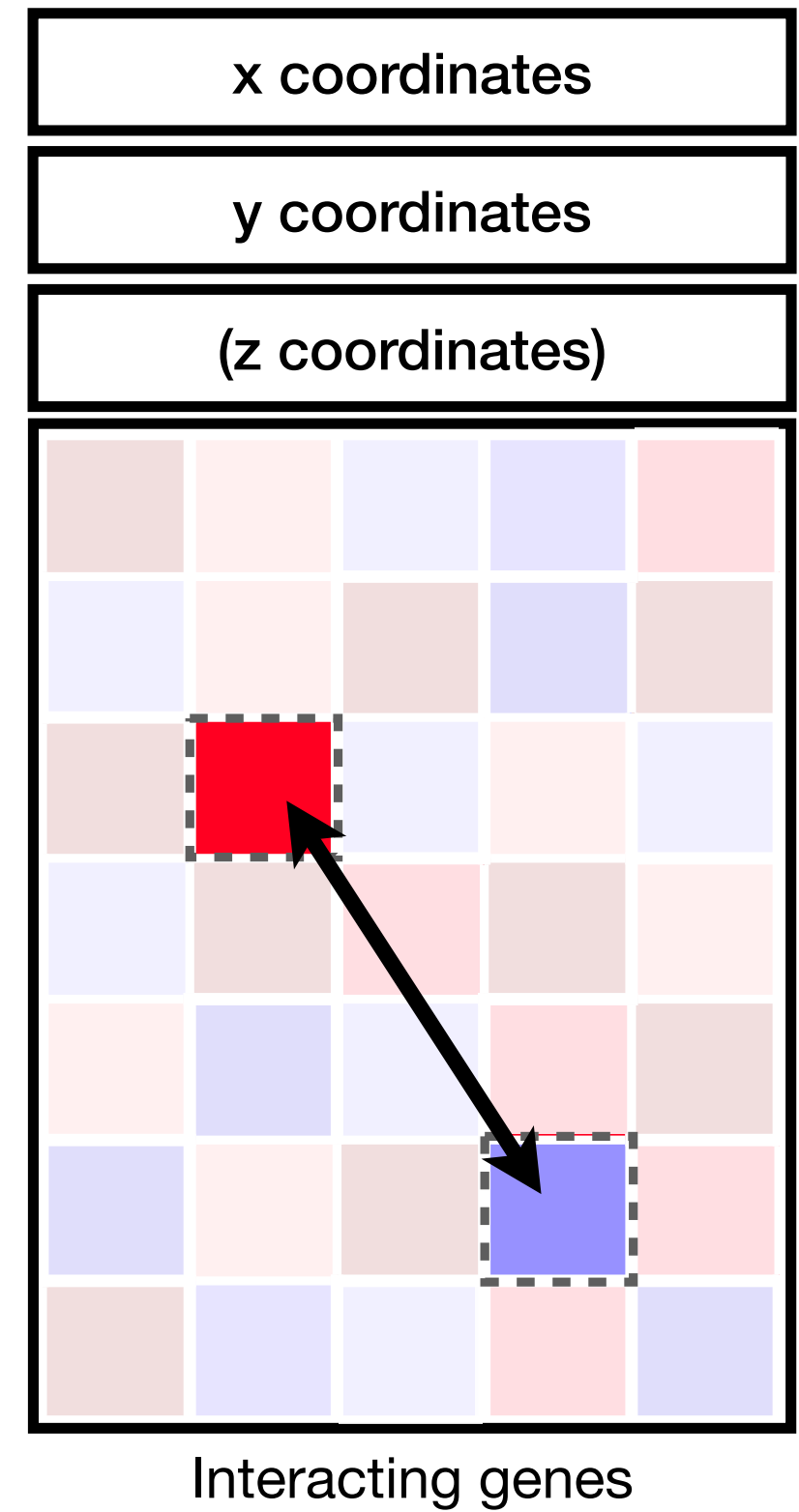

Interacting genes